\documentclass[prb,twocolumn,superscriptaddress,preprintnumbers,amsmath,amssymb,floatfix]{revtex4}
\usepackage{graphicx}

\begin{document} 
\title{Half-metallicity in NiMnSb: \\ a Variational Cluster Approach with ab-initio parameters
}
\author{H. Allmaier}
\email[]{hannes.allmaier@itp.tugraz.at}
\affiliation{Institute of Theoretical Physics, Graz University of Technology,
A-8010 Graz, Austria}
\author{L. Chioncel}
\affiliation{Institute of Theoretical Physics, Graz University of Technology,
A-8010 Graz, Austria}
\affiliation{Faculty of Science, University of Oradea, RO-47800, Romania}
\author{E. Arrigoni}
\affiliation{Institute of Theoretical Physics, Graz University of Technology,
A-8010 Graz, Austria}
\author{M.I. Katsnelson}
\affiliation{Institute for Molecules and Materials, Radboud University of Nijmegen, NL-6525 ED
Nijmegen, Netherlands}
\author{A.I.~Lichtenstein}
\affiliation{Institute of Theoretical Physics, University of Hamburg, 20355 Hamburg, Germany}
 
\begin{abstract} 
Electron correlation effects in the half-metallic ferromagnet 
NiMnSb are investigated within a combined density functional and many-body 
approach. Starting from a realistic multi-orbital Hubbard-model including 
{\it Mn} and {\it Ni-d} orbitals, the many-body problem is addressed via
the Variational Cluster Approach. The density of states obtained in
the calculation shows a strong spectral weight transfer towards the Fermi 
level in the occupied conducting majority spin channel with respect to the 
uncorrelated case, as well as states with vanishing quasiparticle weight
in the minority spin gap. Although the two features produce competing effects,
the overall outcome is a strong reduction of the spin polarisation at
the Fermi level with respect to the uncorrelated case. This result emphasizes 
the importance of correlation in this material.  
\end{abstract} 

\maketitle 

\section{Introduction}

More than twenty years ago {\it de Groot et al.}~\cite{gr.mu.83} 
carried out electronic structure calculations for the half-Heusler compound 
NiMnSb which showed peculiar magnetic features leading to the discovery of
a new class of materials, the so-called half-metallic ferromagnets. Such materials differ 
from conventional ferromagnets in that they display a gap in one of
the spin channels only. The concept of half metallicity boosted the research in 
spintronics - an emergent technology which makes use of spin and charge of electrons at 
the same time. Spintronic applications such as spin-valves, polarized electron 
injectors/detectors or devices using tunneling and giant magneto-resistance effects 
promise to revolutionize microelectronics once highly polarized electrons can be injected 
efficiently at room temperatures~\cite{so.by.98,zu.fa.04}.

Unfortunately, the theoretically predicted ideal full spin polarization of half-metals 
has not yet been found experimentally. As a matter of facts, experiments show that full 
polarization is lost at temperatures of the order of room temperature, and even at lower
temperatures different factors such as structural
inhomogeneities, as well as surfaces and interface properties 
may suppress it~\cite{wi.gr.01,ka.ir.08}.

Despite the fact that high quality NiMnSb films have been successfully
grown, they were not found to 
reproduce the half-metallic character of the bulk suggested by spin-polarized 
positron-annihilation~\cite{ha.mi.86,ha.mi.90}. Values of spin polarization were reported 
between $40\%$ in spin-resolved photoemission measurements~\cite{zh.si.01} up to $58\pm2.3\%$ 
by superconducting point contact measurements at low temperatures~\cite{so.by.98} 
(see also Refs.~\onlinecite{co.ei.06,mi.br.09}). The discrepancy between theoretical 
calculations~\cite{gr.mu.83} and the above mentioned experimental facts were attributed 
to surface and interface effects. 
Consequently, different surface and interfaces of NiMnSb were theoretically investigated 
by {\it de Wijs} and {\it de Groot}~\cite{wi.gr.01}, which demonstrated that half-metallicity 
can be preserved at the surface and/or interface by suitable reconstruction~\cite{wi.gr.01}.
The theoretical situation is
complicated by the fact that the spin polarisation (or, more precisely
the tunneling magnetoresistance)  displays a
substantial uniaxial anisotropy in this material.~\cite{li.gi.06}
 
Recently, finite-temperature correlation effects were addressed in several 
half-metals~\cite{ka.ir.08,ch.ka.03,ch.ma.06,ch.al.07,ch.sa.08,ch.ar.09}. For NiMnSb, a Local Density 
Approximation plus Dynamical Mean Field Theory calculation (LDA+DMFT)~\cite{ch.ka.03} 
showed the appearance of so-called non-quasiparticle (NQP)
states. These states originate from spin-polaron
processes~\cite{ed.he.73,ir.ka.90}, whereby the spin-down low-energy electron excitations, 
which are forbidden for half-metallic ferromagnets in the one-particle picture, turn 
out to be possible as superpositions of spin-up electron excitations and virtual 
magnons~\cite{ed.he.73,ir.ka.90,ka.ir.08}. 
Here, we extend this study by adopting the Variational Cluster Approach (VCA), which
includes correlations beyond the locality captured by DMFT.
In addition, the VCA is based on exact diagonalisation, which is
more appropriate than the diagrammatic method (FLEX) adopted
in Ref.~\onlinecite{ch.ka.03} to solve the impurity problem.
In a previous paper,~\cite{al.ch.08.sp} we used the VCA
to investigate the spin polarization in NiMnSb taking into account only the {\it Mn-d} orbital 
basis set. Our calculations showed that the {\it Mn-d}-only basis set
is not sufficient to appropriately describe 
the low energy spectrum of NiMnSb around the Fermi level. For this reason,
in the present work we adopt a multi-orbital Hubbard-type Hamiltonian
which includes all $10$ 
{\it Mn} and {\it Ni-d} orbitals. Our present calculation confirms that the inclusion of  
the latter is essential for a proper description of ferromagnetic
properties and of the minority spin
gap in NiMnSb.

Our results support the existence of 
states within the minority spin gap in agreement with previous 
LDA+DMFT calculations~\cite{ch.ka.03,ka.ir.08}. In addition, 
they indicate that these so-called nonquasiparticle states indeed have a vanishing
quasiparticle weight at the Fermi energy.
At the same time, our results predict  a correlation-induced 
spectral weight transfer for the majority spin states.  The
combination of these two effects yields a polarization 
whose energy dependence is in qualitative agreement with experiments.
These calculations lead to the conclusion that even in the presence of medium-size interactions, 
electron correlations significantly affect the spin polarisation in half-metals.

This paper is organized as follows: in section~\ref{sec:nmto} we present 
the methods used to investigate the electronic structure of NiMnSb.
In particular, in Sec.~\ref{abin} we describe the {\it ab-initio} construction of the many-body 
model Hamiltonian. Specifically, the uncorrelated part of the Hamiltonian for 
excitations in the vicinity of the Fermi level is obtained from the so-called 
downfolding technique~\cite{an.sa.00,zu.je.05} within the Nth-order muffin tin 
orbital (NMTO) method. In Sec.~\ref{vca}, we give a short summary of the VCA 
approach. We present and discuss our results in Sec.~\ref{resu}. In particular,
in Sec.~\ref{dos}, we evaluate the density of states within VCA and 
discuss the results in the framework of previous calculations.
In Sec.~\ref{spec} we discuss $k$-dependent spectral properties, namely the spectral 
function and the self-energy. Finally, spin polarization and its
comparison with experiments is discussed in Sec.~\ref{polarization},
and the summary of the results is presented in Sec.~\ref{summ}.

\section{Electronic structure calculations for NiMnSb}
\label{sec:nmto}

\begin{figure}[h]
\includegraphics[width=0.9\columnwidth]{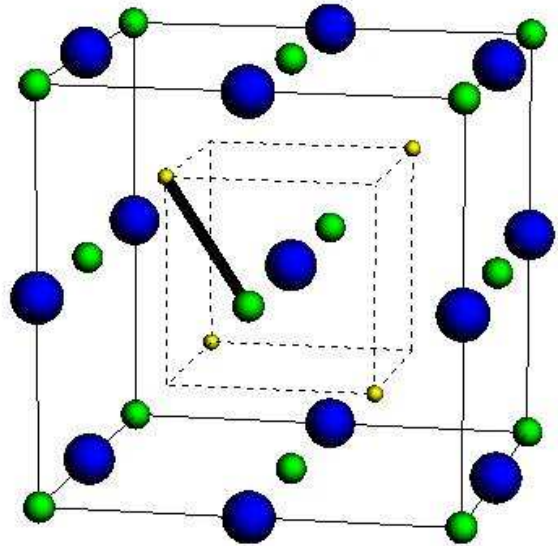}
\caption{(Color online) The conventional unit  cell for the
semi-Heusler NiMnSb compound: {\it Sb} (large, blue spheres) and
{\it Mn} (medium, green spheres) sit on the same faces of the large
cube (shown with thin, solid lines), {\it Ni} (small, yellow
spheres) forms a separate (small) cube drawn using dashed lines.
In addition, the atoms belonging to the reference 
system of the VCA calculation are connected by a thick solid line. 
{\it Sb} orbitals are downfolded and, therefore, not included directly in the model calculation.
}
\label{nimnstructure}
\end{figure}

The intermetallic compound {\it NiMnSb} 
crystallizes in the cubic structure of {\it MgAgAs} type (C1$_b$) with the $fcc$ Bravais lattice 
(space group $F\overline{4}3m=T_d^2$). This structure can be described as three 
interpenetrating {\it fcc} lattices of {\it Ni, Mn} and {\it Sb} with the lattice
parameter  $a=11.20 a_0$ ($a_0=$ Bohr radius), respectively. The {\it Ni} and 
{\it Sb} sublattices are shifted relative to the {\it Mn} sublattice by a quarter of 
the $[111]$ diagonal in opposite directions, see also Fig.~\ref{nimnstructure}. 
The important
aspects~\cite{gr.mu.83,og.ra.95,ga.de.02,na.da.03,ku.ma.90,ka.ir.08} 
which determine the 
behavior of electrons near the Fermi level, as well as the half-metallic properties 
are the interplay between the crystal structure, the valence electron count, the 
covalent bonding, and the large exchange splitting of {\it Mn-d} electrons. 
For the minority spin gap opening, not only the {\it Mn-d}--{\it Sb-p} interactions, 
but also {\it Mn-d}--{\it Ni-d} interactions have to be taken into account. In 
addition, the loss of inversion symmetry produced by the C1$_b$ structure (the 
symmetry lowering from $O_h$ in the L2$_1$ structure to $T_d$ in the C1$_b$ structure)
are important for these effects. The existence of $sp$-valent {\it Sb} is 
crucial to provide stability to this compound.

The crystal structure is shown in Fig. \ref{nimnstructure}. The positions occupied by
atoms are represented by spheres. For illustrative purposes, in Fig.~\ref{nimnstructure}
the radii of the spheres were chosen arbitrarily. The actual muffin-tin radii used in the
calculations are $R_{MT}^{Ni}=2.584; R_{MT}^{Mn}=2.840; R_{MT}^{Sb}=2.981$ and
$R_{MT}^{E}=2.583$ (atomic units) for the vacant position situated in
$(1/4, 1/4, 1/4)$. The LMTO-ASA basis used for the self-consistent
calculations contains the {\it spd}-partial waves for {\it Mn} and {\it Ni},
the {\it sp(df)}-partial waves for {\it Sb} and {\it s(pd)}-partial waves for the
empty sphere {\it E}. ({\it l}) means that the {\it l}-partial waves are downfolded within the
selfconsistent calculations.

\subsection{Ab-initio construction of the model Hamiltonian}
\label{abin}

In order to construct the effective low-energy Hamiltonian to use in our VCA calculation, 
we employed the Nth order muffin-tin-orbitals scheme within the LMTO-ASA basis set~\cite{an.je.84}.
The NMTO method~\cite{an.sa.00,zu.je.05} can be used to generate truly minimal basis sets 
with a massive downfolding technique. Downfolding produces minimal bands which follow 
exactly the bands obtained with the full basis set. The minimal set of symmetrically 
orthonormalized NMTOs is a set of Wannier functions. In the construction of the NMTO 
basis set the active channels are forced to be localized onto the eigenchannel ${\bf R}lm$,
making the NMTO basis set  strongly localized.

Fourier-transformation of the orthonormalized NMTO Hamiltonian, 
$H^{\rm LDA}({\bf k})$, yields on-site energies and hopping integrals,
\begin{equation}
\left\langle \chi^{\perp,A} _{{\bf R'}\,m'}\left\vert H
^{LDA}
-\varepsilon _{F}
\right\vert \chi^{\perp,B} _{{\bf R} \,m}\right\rangle 
\equiv t_{m^{\prime },m}^{A-B,\bf{R'-R}\rm} ,
\end{equation}
in a Wannier representation, where the NMTO Wannier functions
$\left\vert \chi _{\mathbf{R} \,m}^{\perp,A}\right\rangle$ are orthonormal.
Here, $t_{m^{\prime },m}^{A-B,{\bf R'-R}}$ denotes
the hopping term from orbital $m$ of atom $B$ 
on site ${\bf R}$ to the orbital 
$m^{\prime }$ of atom $A$ on site ${\bf R'}$
($A$ and $B$ are either {\it Ni} or {\it Mn}) 
Further information concerning technical details of the calculation can be 
found in Ref.~\onlinecite{ya.ch.06,ka.ir.08}.

In a previous paper~\cite{ya.ch.06} we discussed the chemical bonding and computed model Hamiltonian 
parameters for the semi-heusler NiMnSb using only {\it Mn-d} Wannier orbitals.
As mentioned above, not only the {\it Mn-d}--{\it Sb-p}, but also {\it Mn-d}--{\it Ni-d} 
interactions are required to open a gap in the minoriy spin channel: the minority 
occupied bonding states are mainly of {\it Ni-d} character, while the unoccupied 
anti-bonding states are mainly of {\it Mn-d} character. 
Therefore, in the present work we consider an enlarged NMTO-basis 
consisting of {\it Ni-} and {\it Mn-d} 
orbitals which span an energy window of about $\pm 3eV$ around
the Fermi energy.

The matrix elements for the on-site energies
$\epsilon_{m}^A\equiv t_{m,m}^{A-A,{\bf 0}}$
are given by
(we use the convention in which $m=1,\cdots, 5$ corresponds to
the $d$ orbitals $\{ xy,yz,zx,3z^2{\rm -}1,x^2{\rm -}y^2 \}$ in the order)
%
\begin{equation}
\epsilon_{m}^{\text{Mn}} = \left(-1411,-1411,-1411,-721,-721\right),
\end{equation}
\begin{equation}
\epsilon_{m}^{\text{Ni}} =
\left(-2439,-2439,-2439,-2679,-2679\right)\;.
\end{equation}
The nearest-neighbour hopping terms are given by
(${\bf \Delta_1} \equiv (-\frac{1}{4},-\frac{1}{4},-\frac{1}{4})$)
\begin{equation} \nonumber
 t_{m',m}^{\text{Ni-Mn},{\bf \Delta_1}} = \left(
 \begin{array}{rrrrr}
  -153 & -272 & -272 & -153 &    0 \\
  -272 & -153 & -272 &   76 & -132 \\
  -272 & -272 & -153 &   76 &    0 \\
   110 & - 55 & - 55 &    1 &  132 \\
     0 &   95 & - 95 &    0 &    1
\end{array}
\right)\; ,
\label{Heff_NiMnSb_NiMn_1nn}
\end{equation}
and the next-nearest-neighbour terms
(${\bf \Delta_2} \equiv (\frac{1}{2},-\frac{1}{2},0)$)
\begin{equation}
t_{m',m}^{\text{Mn-Mn},{\bf \Delta_2}} = \left(
\begin{array}{rrrrr}
 -107 &  -14 &   14 &   72 &    0 \\
   14 &    6 &   36 &  -12 &    4 \\
  -14 &   36 &    6 &   12 &    0 \\
   72 &   12 &  -12 &   61 &    4 \\
    0 &   -4 &   -4 &    0 &  -52
\end{array}
\right)\;,
\label{Heff_NiMnSb_2nn_MnMn}
\end{equation}


\begin{equation}
t_{m',m}^{\text{Ni-Ni},{\bf \Delta_2}} = \left(
\begin{array}{rrrrr}
  142 &  -53 &   53 &  129 &    0 \\
   53 &  229 &  -71 &  133 &  -92 \\
  -53 &  -71 &  229 & -133 &  -92 \\
  129 & -133 &  133 &   40 &    0 \\
    0 &   92 &   92 &    0 &  -51
\end{array}
\right)\;.
\label{Heff_NiMnSb_2nn_NiNi}
\end{equation}
Here, all hoppings are given in units of meV, and only one representative 
hopping integral is shown for each class. Other hopping terms can be derived 
from proper unitary transformation using crystal symmetry (see, e.g.,  
Ref.~\onlinecite{pa.ya.05} for details). As one can see, the largest hoppings 
occur between the Wannier orbitals located on {\it Ni-} and {\it Mn} atoms. 
In addition, there are further hopping terms in the Hamiltonian, which we don't 
show here for simplicity. We have taken into account hoppings up to a range
of $r=2.0 a$. Neglected hoppings are about a factor 
30
smaller than the largest nearest-neighbor hopping. The non-interacting part of 
the effective Hamiltonian for NiMnSb, thus, has the form
\begin{equation}\label{h0}
H_0 = \sum_{ {\bf R'},{\bf R},\sigma} \sum_{\{A,B,m',m\}} 
t_{m', m}^{A-B,\bf R'-R} c^{\dag}_{B {\bf R'} m'\sigma} c_{A{\bf R} m\sigma } .
\end{equation}
To take into account correlation effects, we add the spin-rotation invariant interaction $H_I$
\begin{equation}
H_I 
=
 \frac{1}{2}\sum_{{\bf R},A, \sigma,\sigma'} 
\sum_{{m,n,o,p}} U_{mnop} c^{\dag}_{A{\bf R}m\sigma}c^{\dag}_{A{\bf R}n\sigma '} 
c_{A{\bf R}p\sigma'} c_{A{\bf R}o\sigma }
\label{hi}
\end{equation}

In Eqs.~\eqref{h0} and ~\eqref{hi}, $c_{A{\bf R}m\sigma}$ $(c^{\dag}_{A{\bf R}m\sigma})$ are the usual 
fermionic annihilation (creation) operators acting on an electron with spin $\sigma$ at site 
{\bf R} in the orbital $m$ of atom $A$. 
The Hamiltonian we are using includes spin- and pair-flip terms, as especially spin
flip processes are important for a correct description of non-quasiparticle
states~\cite{ka.ir.08,ch.al.07}.  For a realistic description of Coulomb interactions, 
the  matrix elements $U_{mnop}$ can be computed for the particular 
material in terms of effective Slater integrals and Racah or Kanamori coefficients~\cite{im.fu.98,ko.sa.06}.
We used for both {\it Mn} and {\it Ni} the following effective Slater parameters: 
$F_\text{Mn/Ni}^0=1.26$eV, $F_\text{Mn/Ni}^2=5.58$eV and $F_\text{Mn/Ni}^4=3.49$eV, 
which correspond to $U=U_{mmmm}=2.0$eV and give an average $\bar{J}=0.65$eV. In addition, 
we checked that our results do not depend significantly on the chosen $U$ and 
$\bar{J}$-values by performing additional calculations for $U=2.5$eV and $U=3$eV 
with $\bar{J}=0.77$eV. The range of values corresponds to the one used in previous 
works~\cite{ch.ka.03,ch.ka.05,ch.ar.06,ch.ma.06}.

The on-site energies calculated in NMTO already contain effects from the Coulomb interaction 
at the LDA mean-field level. While this double counting can be absorbed into the chemical 
potential when only one set of degenerate orbitals 
is taken into account (see e.g. Refs.~\onlinecite{ch.al.07,al.ch.09}) this is generally not 
possible if the full $d$-orbitals and/or the orbitals of more than one
inequivalent atom are used as basis set. Consequently, in order to avoid 
a double counting of the Coulomb contribution, the corresponding Hartree terms have to be 
subtracted from $H_0$.~\cite{pe.ma.03} This is achieved by replacing
the onsite energies 
$\epsilon_{m}^{A}$ obtained from NMTO with~\cite{paranmto} 
\begin{eqnarray}
\epsilon_{m}^{'A}=&\epsilon_{m}^{A}& - \frac{1}{2}\biggl\{ U_{mmmm} \langle n_{m}^{A}\rangle \nonumber \\
&-& \sum_{m'\neq m}(2 U_{mm'mm'}-U_{mm'm'm})\langle n_{m'}^{A}\rangle\biggr\}\;.
\end{eqnarray}
$\langle n^{A}_{m}\rangle$ denotes the occupation of the {\em NMTO band} associated 
with atom $A$ and orbital $m$. Notice that this is different from the occupation of 
the {\em orbital} $m$ in atom $A$.
The present double-counting procedure corresponds  to the ``around-mean-field'' 
scheme~\cite{cz.sa.94,pe.ma.03}. In order to test the dependence on the double-counting schemes,
we  performed calculations considering different schemes, such as the
``fully-localized'' scheme and a combination of these two schemes as
discussed in Ref.~\onlinecite{pe.ma.03}. From these tests, we find that 
our LDA+VCA results for the spectral function remain unchanged 
for energies within  $E_F\pm 1$eV, while for energies outside this range 
(from $\pm 1$eV to about $\pm 3$eV) a redistribution of {\it Mn-} and 
{\it Ni-} states is obtained.

\subsection{Variational Cluster Approach}
\label{vca}

To solve the many-body Hamiltonian~\eqref{h0}+\eqref{hi},
we employ the Variational Cluster Approach~\cite{po.ai.03,da.ai.04}.
This method is an extension of Cluster Perturbation Theory (CPT)~\cite{gr.va.93,se.pe.00,ov.sa.89}.
In CPT, the original lattice is divided into a set of disconnected clusters, and the 
inter-cluster hopping terms are treated perturbatively. VCA additionally includes ``virtual''
single-particle terms to the cluster Hamiltonian, yielding a so-called reference system, and 
then subtracts these terms perturbatively. The ``optimal'' value for these variational 
parameters is determined in the framework of the Self-energy Functional Approach 
(SFA)~\cite{pott.03,pott.03.se}, by requiring that the SFA grand-canonical potential $\Omega$
be stationary within this set of variational parameters. 
In this work,  we only include the chemical potential of the cluster
as a variational parameter, which
 is necessary in order to obtain a thermodynamically consistent particle density~\cite{ai.ar.05,ai.ar.06}.
 It is not necessary to include a
ferromagnetic field in order to obtain a ferromagnetic phase, since
the symmetry can be broken already at the finite-cluster 
level~\cite{paranmto}.
In this paper, we use a new method, described in Ref.~\onlinecite{lu.ar.09},
to carry out the sum over Matsubara frequencies required in the evaluation of 
$\Omega$, whereby an integral over a contour lying a finite (not small) distance from the 
real axis is carried out.

As a reference system we use a cluster of two 
sites, representing one {\it Mn}-atom and one {\it Ni}-atom (see Fig.~\ref{nimnstructure}),
each having the full-$d$ manifold of five orbitals.
Since we have to consider all five orbitals for each atom, it is
very difficult to use larger clusters, which have to be exactly 
diagonalized many times in combination with the variational procedure.

\section{Results}
\label{resu}
\subsection{Density of states}
\label{dos}

\begin{figure}[h]
\includegraphics[width=0.95\columnwidth]{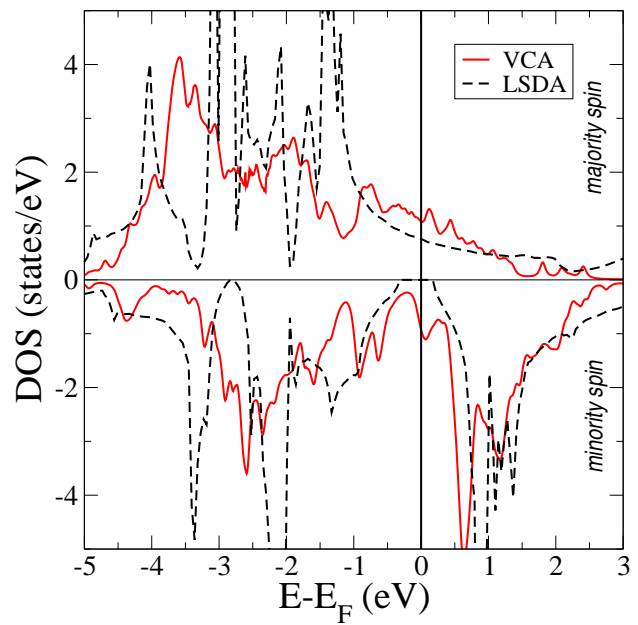}
\caption{(Color online) Density of states  for NiMnSb obtained from
LDA+VCA (red, solid line) for values of the average Coulomb and exchange
parameters $U=2$eV and $\bar{J}=0.65$eV for both Mn and Ni atoms.
in comparison to results obtained from LSDA (black, dashed line).}
\label{dos_vca}
\end{figure}

In order to study the influence of correlations on the half-metallic gap, we first display
the spin-resolved local density of states in Fig.~\ref{dos_vca}. Here, we present a 
 comparison of the results obtained from LSDA with the results from our LDA+VCA calculation. 
The LSDA-DOS is mainly characterized by a large exchange splitting (about $3$eV) of the 
{\it Mn-d} states, leading to large spin moments on the {\it Mn}-site ($3.72\mu_B$). 
A small induced ferromagnetic moment is present on {\it Ni} ($0.29\mu_B$), while the 
{\it Sb} moment ($0.06 \mu_B$) is  anti-parallel to the {\it Mn} moment. Overall,
 the calculated moments are in very good agreement with previous 
ab-initio results~\cite{gr.mu.83,ka.ir.08,ga.de.02,ya.ch.06,ku.ma.90}.    
The existence of large localized {\it Mn} moments of about $3.78\mu_B$
has been verified experimentally by neutron diffraction~\cite{ho.pi.97} as well as
by the sum rule of the x-ray magnetic circular dichroism
spectra~\cite{ki.su.97}. These two experiments also confirm  the magnitude of the 
LSDA-computed moments for {\it Ni} and {\it Sb}. The gap in the minority spin 
channel is about $0.5$eV wide and the total magnetic moment has an integer value of 
$4\mu_B$. Note, that in Fig.~\ref{dos_vca} in LSDA the minority occupied bonding 
states are mainly of {\it Ni-d} character, while unoccupied anti-bonding states are mainly 
of {\it Mn-d} character. It was pointed out~\cite{gr.mu.83} that the opening of a gap 
is assisted by {\it Sb} through the symmetry lowering with the consequence that 
the distinction between {\it Mn-t}$_{2g}$ and {\it Sb-p} character of the 
electrons is lost.  
 
Concerning the LDA+VCA results, we find a total magnetic moment of 3.7$\mu_B$, which
is in reasonable agreement with experimental values~\cite{ho.pi.97,ki.su.97}.
In direct comparison to LSDA, our results show that correlation effects do 
not affect too strongly the general picture of the minority spin DOS 
for energies which are more than 0.5 eV away from the Fermi energy.
 In the range  $0.5 eV\lesssim E-E_F \lesssim 3 eV$, unoccupied {\it Mn-} states 
are visible in the minority spin sector, similarly to LSDA. However, these states  
are shifted to  lower energies due to a slight reduction of the {\it Mn-} exchange 
splitting generated by the many-body correlations. 
Just above the Fermi level,  NQP states 
are present, with a peak around the energy of $0.06$eV. It is important to note that 
these states were also obtained in previous calculations using a LDA+DMFT
many-body approach~\cite{ch.ka.03} at finite temperatures. 
In comparison to the  DMFT description, the non-local correlations captured
by VCA enhance the spectral weight of the NQP states and slightly
shift their position. 
This fact leads to the conclusion 
that even a local DMFT description is sufficient to demonstrate the existence 
of NQP states as discussed previously~\cite{ch.ka.03}. The spectral weight 
of the NQP states is large enough, so that we expect them to be 
well pronounced in  corresponding experimental data. While model calculations 
for single-band Hamiltonians~\cite{ka.ir.08} suggest that NQP states should 
only touch the Fermi level with zero weight at $T=0K$, in our VCA calculation  
they maintain a finite weight at the Fermi level, thus leading to a reduction of spin
polarisation, even at $T=0K$. 
In the LSDA-results, the bonding states below the Fermi level have dominant {\it Ni-d} character
and are responsible for the gap formation. While these states form a single peak at -1.5eV 
in LSDA, the LDA+VCA-results show a splitting into two peaks centered around $-1$eV.
One of these peaks is pushed closer to the Fermi level, while the
other one is shifted to higher energies. The latter correlation effect
is also seen in previous LSDA+DMFT results~\cite{ch.ka.03}.

A significantly stronger effect caused by many-body correlations is visible 
in the majority-spin channel (see Fig.~\ref{dos_vca}). Here we discuss
the behaviour in the same energy range within $\pm 3$eV around the
Fermi level, since this is the 
energy window spanned by our NMTO basis.
The LSDA density of states in 
this energy range is determined mainly by the covalent {\it Ni-Mn-d} 
hybridization, and by the large exchange splitting of {\it Mn-d} 
electrons~\cite{gr.mu.83,ka.ir.08,ya.ch.06}. At the Fermi level and above 
a reduced density of states is present.  
The density of states obtained from the LDA+VCA calculation shows a very strong spectral 
redistribution for the majority spin electrons: the LSDA peak situated around $-3$eV
is lowered in energy while in the energy range between -$2$eV and $E_F$, a spectral-weight
transfer towards the Fermi level takes place. In particular, the large LSDA-peak 
at -1.5eV is shifted to about $-1$eV, which results in a significant contribution 
to the states at the Fermi level. Just above the Fermi level, at energies where NQP 
states are formed in the minority-spin channel, a resonance peak is present for the 
majority-spin electrons. A further maximum of the density of states is present at 0.5eV. 
The meaning of this maximum will become clear in the Sec.~\ref{spec} were the {\bf k}-resolved
spectral functions   are discussed. In contrast to our VCA calculation, DMFT results~\cite{ch.ka.03}
do not change significantly the picture for the majority-spin states. Although
the LDA+DMFT density of states shows a similar reduction of spectral weight
for the peak at $-2$eV, its position remains unchanged. The differences 
between these two results might be explained by the fact that 
within DMFT {\it Mn} and {\it Ni} atoms are only coupled via the general
many-body and charge-self consistency conditions, while correlations are treated 
independently in the two atoms. In contrast, the present VCA approach exactly 
includes correlations on the length scale of the cluster.
These interatom correlations are possibly responsible for the 
splitting of the covalent {\it Ni-Mn-d} electron hybridization in 
the majority spin states. Due to the breaking of this
hybridization, the {\it Mn-d} exchange splitting is decreased, 
which could explain the slight shifts of the minority 
unoccupied and occupied majority {\it Mn-d} states.

\subsection{Spectral properties}
\label{spec}

\begin{figure}[h]
\includegraphics[width=0.9\columnwidth]{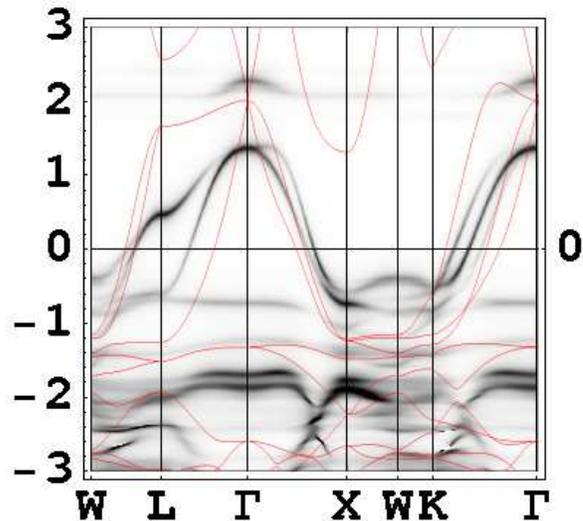}\\
\caption{(Color online)
Majority spin LDA+VCA spectral-function of NiMnSb (black/white density plot) 
along the conventional path in the BZ. 
$W(0.5,1,0)$, to $L(0.5,0.5,0.5)$ through  $\Gamma(0,0,0)$, $X(0,1,0)$,  
$K(0,0.75,0.75)$ points and ending at $\Gamma(0,0,0)$.
The LSDA bands (red, thin solid lines) are shown for comparison. 
Parameters are as in Fig.~\ref{dos_vca}.
\label{spect_vca_up}
}
\end{figure}

\begin{figure}[h]
\includegraphics[width=0.9\columnwidth]{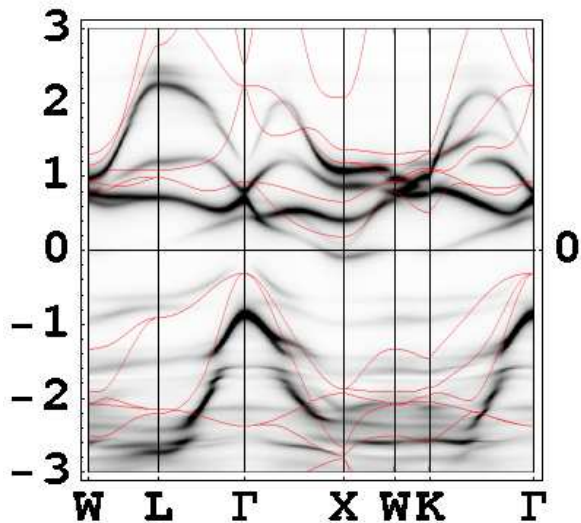}
\caption{(Color online) Minority-spin LDA+VCA spectral function
of NiMnSb (black/white density plot) along the same BZ path as in
Fig.~\ref{spect_vca_up}. The LSDA bands (red, thin solid
lines) are shown for comparison. Parameters are as in Fig.~\ref{dos_vca}.
}
\label{spect_vca_dw}
\end{figure}

In order to gain insight into the nonlocal features of the density of states, 
we compute the {\bf k}-resolved spectral function $A({\bf k}, \omega)$. 
Majority- and minority-spin spectral functions are presented in 
Figs.~\ref{spect_vca_up} and ~\ref{spect_vca_dw}, respectively, with {\bf k}
following high-symmetry points in the Brillouin zone (BZ). The explanation of the 
main features of the LSDA band structure was provided by {\it de Groot et. al.}
in his pioneering paper~\cite{gr.mu.83}. Emphasis was placed on the interaction 
between {\it Mn} and {\it Sb} connected by the symmetry constraint,
while less attention was  
given to the {\it Ni} atom, although {\it Mn} and {\it Ni} are
first-neighbors and a strong 
hybridization between them is evidenced in the density of states. In
our LDA+VCA calculation,  
{\it Ni-d} and {\it Mn-d} states are included explicitly, while 
{\it Sb-} states are admixed  by the downfolding procedure.

Due to correlation, the majority-spin bands crossing the Fermi energy
are substantially narrowed with respect to the uncorrelated LSDA bands.
Specifically, our results show for the bands crossing the Fermi level a 
reduction of the bandwidth from $3.2$eV to $2.2$eV.
Along the path $W \rightarrow L$ both LSDA bands and the VCA 
spectral function cross the Fermi level at almost the same {\bf k}-point.
The degenerate unoccupied level situated in the $L$-symmetry
point, around 1.5eV, is strongly pushed towards the Fermi energy, and 
determines the appearance of the peak visible at $0.5$eV in the DOS 
discussed in Sec.~\ref{dos}. At the same time, correlation effects  
further split the degenerate levels at the $\Gamma$-point seen 
in LSDA at around 2eV. Note that along $\Gamma \rightarrow X$
crossing of the Fermi level occurs close to the corresponding 
crossings of the LSDA-bands.
Furthermore, along the path $X \rightarrow W \rightarrow K$ 
both VCA and LSDA bands are only weakly dispersive. However, the 
VCA bands are shifted towards the Fermi level, while along the line 
back into the $\Gamma$-point, the Fermi-energy crossing of the VCA
bands takes place closer to the $K$-point.

The minority-spin band structure of LSDA shows an indirect gap of about 0.5eV between 
$\Gamma$ and $X$-point. Within this indirect gap formed by the mostly {\it Ni-d} 
occupied and mostly {\it Mn-d} unoccupied states, the LDA+VCA results show 
substantial spectral weight, as can be seen in Fig.~\ref{spect_vca_dw}. 
Notably, across the Fermi level a weakly-dispersive band is present, centered
around  $0.1$eV, representing the NQP states. At higher energies, in the range 
of $1$ to $2$eV above $E_F$, the VCA bands are substantially correlation-narrowed 
with respect to LSDA. The features above the Fermi level, including the 
non-quasiparticle states, have dominant {\it Mn-d} character.
Below the Fermi level, correlations split off the occupied bands having mainly 
{\it Ni-d} character. The spectral weight is redistributed: a part is transfered 
towards the Fermi level, however with smaller weight, while most weight is transfered 
towards higher binding energies. The same effect is visible in 
the density of states plot displayed in Fig.~\ref{dos_vca}. 
Notice that while the shift towards higher binding energies is also seen 
in the previous LSDA+DMFT calculation~\cite{ch.ka.03}, the weak shift 
towards the Fermi level is only obtained within the present calculation.

\begin{figure}[h]
\includegraphics[width=0.9\columnwidth]{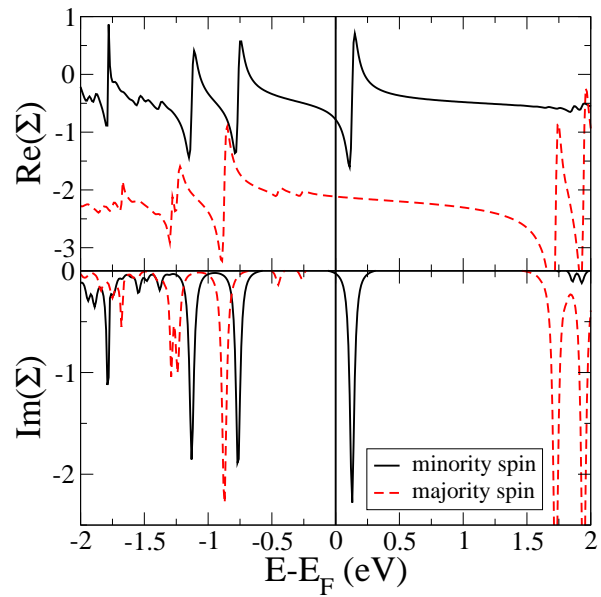}
\caption{(color online) Spin resolved  self-energy on {\it Mn-d} orbitals, at the crossing point $k=(0.5,0.7,0.3)$.}
\label{sigma}
\end{figure}

In order to explore correlation effects in more detail, we plot in
Fig.~\ref{sigma} the self-energy on {\it Mn} sites. 
in the energy range $\pm 2$eV around the Fermi level and near the Fermi crossing at 
$k=(0.5,0.7,0.3)\pi/a$. 
The upper/lower panel of Fig.~\ref{sigma} shows the spin resolved real/imaginary part 
of the electronic {\it Mn} self-energy. 

Below $E_F$, the minority-spin self-energy is similar  to the self-energy of the majority spin channel.
Just above the Fermi level, however, a clear peak in $\text{Im} (\Sigma_{VCA}^{\downarrow})$ is present 
with a maximum around the energies of the non-quasiparticle states (Fig.~\ref{dos_vca} and
Fig.~\ref{spect_vca_dw}).
In previous DMFT calculations~\cite{ch.ka.03}, a very similar 
behaviour of the imaginary part of the local self-energy was seen. In
that case, the pronounced feature above $E_F$ was attributed to the 
minority {\it Mn-d}(t$_{2g}$) states. 
The real part of the self-energy  displays a negative slope 
$\partial \Sigma/\partial \omega<0$ at the Fermi energy for both spin
directions, which confirms that the quasiparticle weight
$Z = (1-\frac{\partial \Sigma}{\partial \omega})^{-1}$ is reduced by correlations.
However, while for majority spins 
$|\partial \Sigma_{\uparrow}/\partial \omega|$ is clearly less than unity, 
for minority spins
$|\partial \Sigma_{\downarrow}/\partial \omega|\gtrsim 1$ (within our
approximation, we cannot determine $\Sigma$ with sufficient accuracy), 
suggesting the nonquasiparticle nature of 
the minority spin states within the gap.

\subsection{Low-energy spin polarization and comparison with experiments}
\label{polarization}

\begin{figure}[h]
\includegraphics[width=0.95\columnwidth]{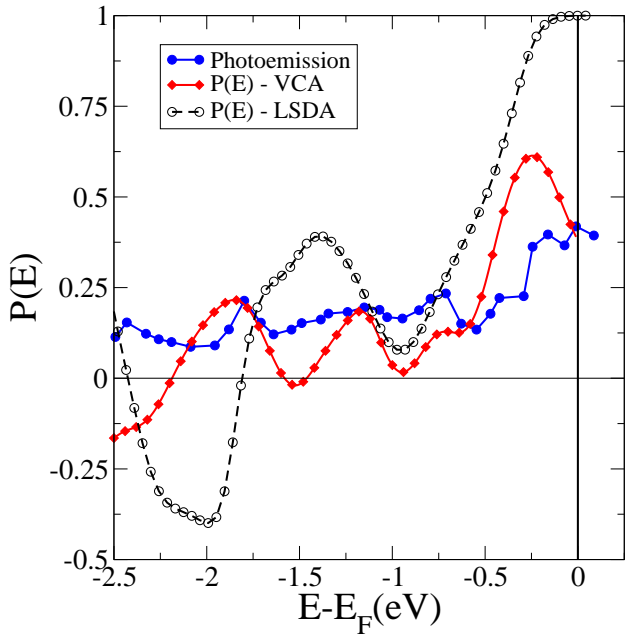}
\caption{(Color online) Energy-dependent polarization obtained from
LSDA and LDA+VCA in comparison with data from spin-polarized photoemission~\cite{zh.si.01}.}
\label{polar}
\end{figure}

To investigate the consequences of the modification of majority and minority
spectral weight at the Fermi energy produced by correlations,
we turn to the issue of the spin polarization.
This is given by the expression
$P(E)=(N_{\uparrow}(E)-N_{\downarrow}(E))/(N_{\uparrow}(E)+N_{\downarrow}(E))$,
 $N_{\sigma}(E)$ being the spin-resolved density of states, 
and is plotted in Fig.~\ref{polar} as a function of energy measured from the Fermi
level. The computed LSDA and VCA values are compared with the raw data obtained from 
spin-resolved photoemission measurement by Zhou et al.~\cite{zh.si.01}. 
For this comparison, the density of states was multiplied with the
Fermi function 
and a Gaussian broadening of 100meV was used to account for experimental resolution.
{\it Zhu et.al}~\cite{zh.si.01} discuss the appearance of a shoulder
close to the
Fermi level when proper annealing
is performed to restore the stoichiometry in NiMnSb. This shoulder is visible in the majority
spin channel (Fig.2c from Ref.~\onlinecite{zh.si.01}) and could be an indication for the 
correlation-induced spectral weight transfer of the majority spin states, not present in the 
LSDA calculations. In addition, the value of the spin polarization at the Fermi level
obtained from our  LDA+VCA-results is situated in the interval of values reported
experimentally~\cite{cl.mi.04,zh.si.01,so.by.98}.

\section{Summary}
\label{summ}

We have investigated the effects of correlations in NiMnSb
using a combined LDA+VCA approach. The parameters for the effective non-interacting 
Hamiltonian were obtained using the downfolding procedure, for a basis including 
{\it Ni} and {\it Mn-d} orbitals. The multi-orbital Hubbard-type many-body 
Hamiltonian was solved using the Variational Cluster Approach for different values 
of $U_{Mn/Ni}$ in the range of $2-3$eV and $\bar{J}_{Mn/Ni}=0.65/0.78$eV. The results presented 
do not show significant differences for the studied range of parameters, nor for 
different double-counting procedures used.
We showed that the presence of {\it Ni-d} orbitals in the NMTO-basis
allows for a more complete description of the low-energy behavior of NiMnSb.
In particular, it correctly describes the spectral weight transfer towards the 
Fermi level in the majority spin channel and the formation of minority-spin 
states with vanishing quasiparticle weight (NQP states) just above the
Fermi level. 
The analysis of the minority-spin spectral function shows for the NQP states a 
weakly dispersive band having dominantly {\it Mn-d} character.
Due to electron correlations, the covalent {\it Ni-Mn} d-hybridization in the 
majority-spin channel splits up and part of the weight is transferred
towards the Fermi level.
The {\it simultaneous} presence of majority spin spectral weight transfer towards
the Fermi level, and the occurrence of minority-spin non-quasiparticle states 
emphasizes the importance of correlation effects in this material, despite the 
small value of $U$.

Despite the fact that high-quality films of NiMnSb have been grown,
they do not reproduce the half-metallic character of the bulk detected by spin-polarized 
positron-annihilation~\cite{ha.mi.86,ha.mi.90}. On the other hand, one should mention that
the positron annihilation technique only provides an evidence for half
metallicity by means of a consistency check. In other words, the ``proof'' is carried out 
by modeling the data assuming a half-metallic band structure, with a full minority spin
gap, {\em from the outset}~\cite{ha.mi.86,ha.mi.90}. For this reason, it would be interesting 
to revisit the analysis of the positron-annihilation data
by using the correlated band structure 
obtained here, i. e., by
taking into account the existence of NQP states. 

We acknowledge financial support by the Austrian science 
fund (FWF project P18505-N16), 
and by the  cooperation project ``NAWI Graz''  (F-NW-515-GASS).   

\bibliographystyle{prsty}
\bibliography{references_database,footnotes}

\begin{thebibliography}{10}
\bibitem{paranmto}
Notice that the initial NMTO calculation is carried out in the nonmagnetic
  phase, so that the starting Hamiltonian \eqref{h0} plus \eqref{hi} is spin
  independent. The ferromagnetic solution is then obtained variationally within
  the VCA approach. This is important in order to preserve a spin-rotation
  symmetric model Hamiltonian.
\end{thebibliography}

\end{document}